# Growth of solid conical structures during multistage drying of sessile poly(ethylene oxide) droplets

David Willmer[a], Kyle Anthony Baldwin[a], Charles Kwartnik[a,b], David John Fairhurst[*a]

[a]Nottingham Trent University, Clifton Lane, Nottingham, NG11 8NS, UK.
[b]University Institute for Technology Limousin, 87065, Limoges, France

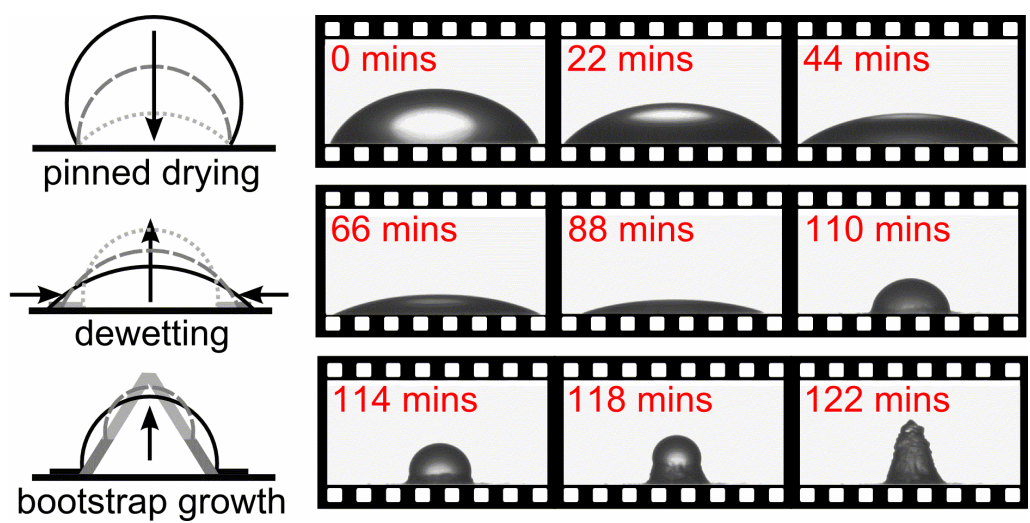

**Microlitre polymer droplets deposit solid conical structures, via a novel bootstrap drying mechanism, over a range of initial conditions.**

## Abstract

Sessile droplets of aqueous poly(ethylene oxide) solution, with average molecular weight of 100kDa, are monitored during evporative drying at ambient conditions over a range of initial concentrations $c_0$. For all droplets with $c_0 \geq 3\%$, central conical structures, which can be hollow and nearly 50% taller than the initial droplet, are formed during a growth stage. Although the formation of superficially similar structures has been explained for glass-forming polymers using a skin-buckling model which predicts the droplet to have constant surface area during the growth stage (L. Pauchard and C. Allain, *Europhys. Lett.,* 2003, **62**, 897–903), we demonstrate that this model is not applicable here as the surface area is shown to increase during growth for all $c_0$. We interpret our experimental data using a proposed drying and deposition process comprising the four stages: pinned drying; receding contact line; "bootstrap" growth, during which the liquid droplet is lifted upon freshly-precipitated solid; and late drying. Additional predictions of our model, including a criterion for predicting whether a conical structure will form, compare favourably with observations. We discuss how the specific chemical and physical properties of PEO, in particular its amphiphilic nature, its tendency to form crystalline spherulites rather than an amorphous glass at high concentrations and its anomalous surface tension values for $M_W = 100$ kDa may be critical to the observed drying process.

## Introduction

The behaviour of complex fluids under non-equilibrium conditions is of everyday relevance in hair washing (dilution and shear of surfactants)[1, 2], food preparation (temperature changes, shearing and diluting of emulsions etc.)[3, 4] and ink-jet printing (drying of colloidal suspensions)[5, 6], to list just three examples. Investigation of such processes from a fundamental perspective may lead to products with greater functionality, improved efficiency, lower costs or reduced environmental impact.

When the drying liquid is a complex fluid containing mixtures or suspensions, the behaviour can be complicated, so various model experimental systems are used. Deegan et.al[7] investigated the formation of the familiar two-dimensional coffee-ring stain using a model system of very dilute micro-spheres suspended in water. They concluded that enhanced evaporation along the pinned contact line, due to a contact angle less than 90°, must be fed by outward flow of water from the centre of the droplet. Suspended particles, such as coffee grains, are carried to the periphery in the flow and deposited at the edge leading to the ring-like pattern. Recently, Hu and Larson showed that ring-formation can be disrupted in the presence of recirculating currents caused by Marangoni flow[8]. Parisse and Allain investigated the changing profile of droplets of concentrated suspensions as they dry[9], observing a gelled three dimensional deposit ("foot") near the drop edge which progressively grows inwards. Alain and Pauchard[10] used the model system of the branched aqueous polymer dextran to investigate the additional complexities that arise as polymer solutions evaporate. In this case, the increase in polymer concentration at the droplet's edge, due again to the outward flux of water, resulted in a phase change: on the surface of the liquid droplet a glassy skin with spherical cap geometry formed which was flexible and permeable, but also incompressible. Further evaporation of water within the droplet led to the glassy cap deforming and buckling, the various shapes of which have been analysed theoretically[11]. Another model system is that of a mixture of a hydrophobic and a hydrophilic liquid, investigated by Rowan *et.al.*[12]. These droplets initially dried to a flat puddle with a contact line that was pinned but that rapidly retreated later causing a nearly spherical droplet to "ball up" from the puddle – an effect driven by an increase in the surface tension as the hydrophilic component evaporated first, thus increasing the contact angle of the solution.

Important to understanding all these observations, and also to this work, is the interplay of surface forces which lead to the observed contact angle $\theta$, measured at the contact line between the tangent to a droplet's surface and the substrate. Ideally $\theta$ is determined uniquely by the balance of the three pairwise surface tensions between solid, liquid and gas regions respectively. (The same result can also be found by consideration of surface energies.) These surface tension values are affected by the nature of the solid surface and the concentration of the solution. In practice however, the contact angle can cover a range: the minimum value, just before the contact line retreats towards the liquid is called the receding contact angle; the maximum $\theta$, as the contact line starts to expand away from the liquid is the advancing contact angle; between these extremes, the contact line remains stationary. As often happens, the contact line can become pinned to microscopic or molecular defects on the substrate leading to a receding contact angle of only a few degrees. The pinned drying scenario leads to outward flows within the droplet.

In this work we investigated the drying of poly(ethylene oxide) (PEO) solutions which, unlike dextran, is a linear (non-branched) polymer and does not exhibit a glass transition but rather precipitates a solid phase (usually as semi-crystalline spherulites) at high concentrations. We were particular interested to see which behaviour PEO would exhibit during drying, forming a buckled skin like dextran, or with pinned drying and ring-stain formation as seen in particle suspensions.

## Experimental method

Solutions were prepared using polymer with an average molecular weight $M_w \approx$ 100,000 (Sigma Aldrich 181986) and calculated radius of gyration[13] $r_g = 10$nm giving an overlap concentration $c^* \approx 4\%$ wt. Solutions spanning a range of initial concentrations $c_0$ from 1% to 45% by mass were mixed by hand using distilled, de-ionised water and were left to equilibrate for at least 24 hours before use. Mechanical mixing methods were avoided (vortex mixer, centrifuge or sonicator) to prevent possible damage to the polymer. The solutions, in particular at the higher concentrations, appeared slightly cloudy due to small undissolved clusters, which can be removed with filtration. However, as the clusters do not seem to affect the nature of the drying, the results presented here are all performed with unfiltered samples.

For each measurement, a droplet with V ~ 50µL was placed onto a glass microscope slide, first cleaned with isopropanol to remove dust and grease. The droplet was dispensed over several seconds using a Hamilton 710 microlitre syringe through a 0.2mm radius needle. Despite the large shear rates in the needle (~ $100s^{-1}$) no significant differences in drying behaviour was seen when compared with droplet deposited less controllably by pouring, so we assume the polymer molecules are undamaged. The droplet was then left to evaporate in an observation chamber (measuring 0.6m by 0.75m by 0.94m) at ambient conditions where the temperature was monitored to within 0.5°C. The chamber was sufficiently large that droplet evaporation did not change the humidity of the environment. A digital camera and diffuse light source (from Krüss) placed either side of the droplet in the chamber were used to record the drying process. Care was taken to place the slide horizontally and to reduce convective air currents around the droplet due to the light source; two effects which can interfere with the deposition process. Digital images of the drying droplet were recorded at 10 second intervals and analysed using Krüss Drop Shape Analysis software. At early times when the droplet surface was smooth, the profile was fitted using the Young-Laplace equation[14] and values for the droplet base radius r, height h, volume V, surface area A and contact angle $\theta$ were extracted. However, once deposition began and the liquid droplet was resting on solid deposit, the Young-Laplace equation could no longer be used to model the entire surface. Instead, the two dimenstional droplet profile was extracted from the recorded images using ImageJ software (from US National Institutes of Health) and the surface area A and volume V of rotation calculated numerically in Matlab using the maximum point on the profile to define the vertical axis of rotation. Uncertainties in V and A due to droplet asymmetry were quantified by halving the difference between the contributions from the profile on either side of the rotation axis, and are very sensitive to variations in the position of this axis, caused by changes in the maximum point.

## Results

Figure 1 shows time-sequence images for four values of $c_0$, indicating the drying stages (discussed below) and the shape of the final deposits, which vary repeatably with $c_0$. Videos of the process are also available in the Electronic Supplementary Information[†] (Videos 1-4). Low concentration droplets ($c_0 < 3\%$) leave a disk-like deposit of solid PEO with diameter equal to the initial droplet base radius $r_0$ and thickness from surface profilometry[15] of around 100µm. Changes in colour and optical transmission indicate that the polymer concentration varies across the disk, suggesting that pinned drying and ring-stain formation occur at these concentrations. For $c_0 \geq 3\%$, in addition to the thin disk, there is a solid deposit usually at or near the centre of the disk. Below 12% the deposit is several millimetres in height, with steep rough surfaces. For $c_0 > 12\%$ the deposit is smoother and conical, the edge extending almost to $r_0$ for $c_0 = 45\%$, with the thin disk continuing beyond. In some "failed" experiments, the deposit falls over during formation, due usually to an inclined substrate or air currents within the chamber. Data from such experiments are not included in the subsequent analysis and discussion. For all $c_0 \geq 3\%$, the final structure is a rough white deposit, which, when viewed from underneath, often shows a hollow region in the very centre adjacent to the glass coverslip.

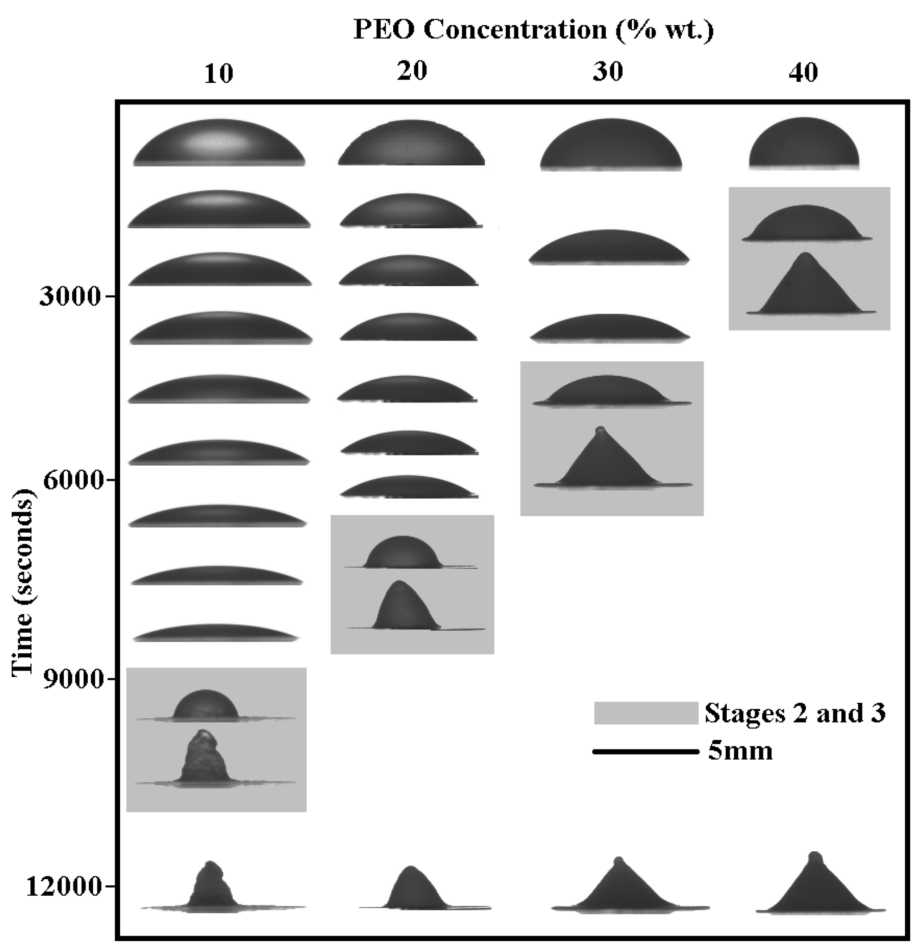

**Fig.1** Snapshots illustrating the drying process for droplets at four values of $c_0$. The grey boxes indicate stages 2 and 3, during which the droplet height increases. Videos can be found in the web supplement[†].

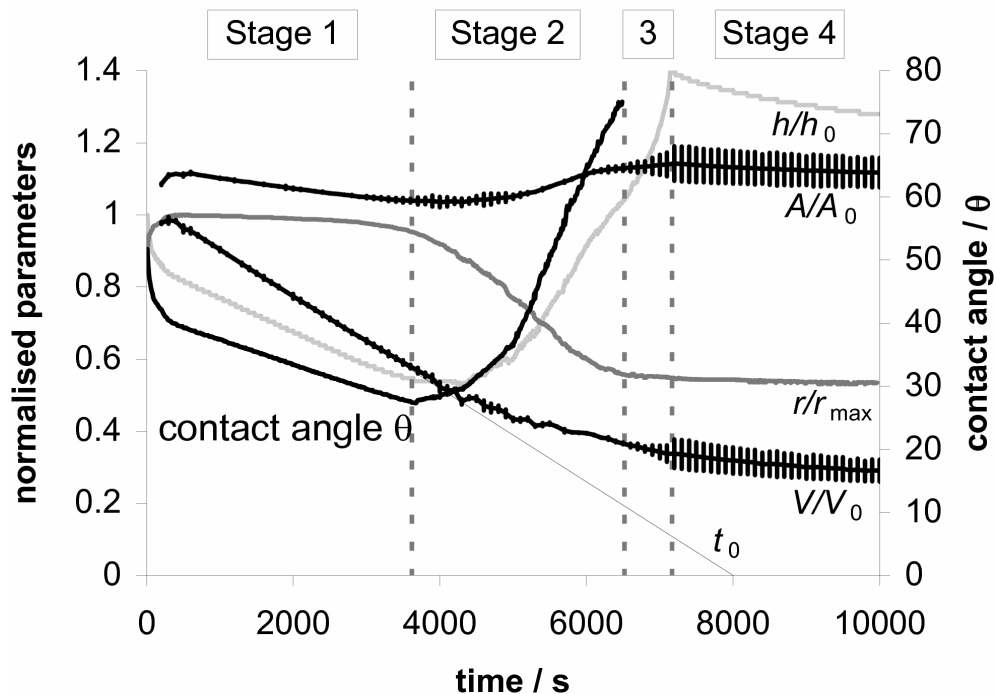

**Fig. 2** Measured normalised properties of a drying droplet with $c_0 = 25\%$ as a function of time. Values of contact angle $\theta$ are obtained using Krüss drop shape analysis software to fit the Young-Laplace equation to the droplet profile, and become meaningless when the droplet is no longer only liquid. Values of height h, surface area A, base radius r and volume V are calculated by numerically integrating digitised droplet profiles. The linear extrapolation of V is shown as a thin line, and the intercept on the t-axis gives $t_0$. Uncertainties in A and V are due to asymmetric droplets.

Fig.2 shows values of V, A, r, h and $\theta$ extracted from the recorded images of a drying droplet with $c_0 = 25\%$. The values V, A and h are normalised by their initial values, $V_0$, $A_0$ and $h_0$ respectively, and r by its maximum value $r_{max}$. Before deposition begins (around t = 4000s in Fig 2), values calculated using our routine to numerically integrate the droplet profile were indistinguishable from those determined using the commercial DSA software which fits the Young-Laplace equation, so we plot only the values from numerical integration, as these are also reliable after deposition has begun. Within the first few minutes, the droplet spreads slightly as seen by an increase in r and A and a decrease in h and $\theta$. For the following hour the droplet loses volume linearly, while the contact line is pinned so r remains constant as h and A decrease. As in other work[10], we extrapolate the linear portion of V to intercept the time axis and use this value to define the time it would take the droplet to dry to zero volume, $t_0$:

$$t_0 = -\frac{V_0}{\left(\dfrac{\partial V}{\partial t}\right)_{t=0}} \qquad (1)$$

Experimental times can be normalised by $t_0$ to compensate for uncontrolled variations in initial droplet volume and relative humidity of the chamber. During this initial period when the volume loss is constant and A decreasing, the average evaporative flux across the interface must be increasing, which is predicted to occur as $\theta$ decreases[7, 10].

After just over an hour (t=3620s) h reaches its lowest value $h_{min}$ at time $t_{min}$. From this point the contact line begins to contract (r decreases) causing h, A and $\theta$ to increase, while the volume continues to reduce, albeit, at a slower rate than initially. Around fifty minutes later (t=6520s) significant deposition in the centre begins so values of $\theta$ from DSA processing become meaningless, r remains constant, V continues to decrease and A and h continue to increase, with h slowing down and then accelerating again. After another ten minutes (t=7170s) h reaches its maximum value $h_{max}$ at time $t_{max}$. The deposit then contracts slowly for up to three more hours until changes become imperceptible, but we chose to omit this late stage data from Fig. 2.

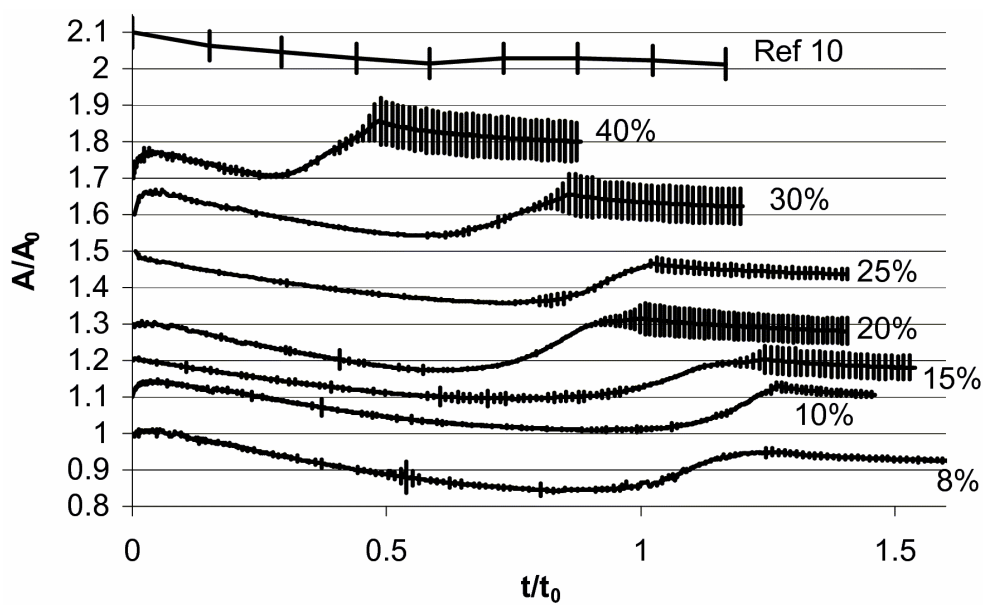

**Fig.3** Normalised surface area of drying droplets calculated using numerical integration of droplet profiles extracted from digital images, plotted against normalised time $t/t_0$. For clarity, curves are offset vertically. Error bars represent uncertainties due to asymmetric droplets, in particular at later times when the highest point of the profile (and therefore the axis of rotation) is no longer central. Occasionally the uncertainties increase (e.g. on 8% curve close to $t/t_0 = 0.5$) due to bright reflections from the top of the droplet leading to a cusped profile and a peak position which jumps around. For comparison, the upper curve shows data for the skin buckling model with dextran[10] in which the surface area remains constant during the growth phase.

In Fig.3 we compare the evolution of $A/A_0$ for seven representative values of $c_0$ alongside literature data for dextran[10], with error bars determined as described above. The short bursts of high uncertainty in the early part of several of the curves are caused by the illuminating light reflecting from the top of the droplet and confusing the image processing routine which then finds a profile with a slight cusp. This cusp leads to the maximum point on the profile jumping horizontally a few pixels, giving noticeable differences in A between the two sides and therefore large uncertainties. As the droplet dries further, the reflected light no longer appears at the top of the droplet so the image processing routine extracts a correct profile and the errors reduce again. At later times, the uncertainties are due to the true asymmetrical shape of the deposit. We see a very early increase in surface area as the droplet spreads, followed by a period in which A steadily decreases. Even within our realistic error bars, all concentrations show a significant increase in surface area during the time when the droplet height is increasing.

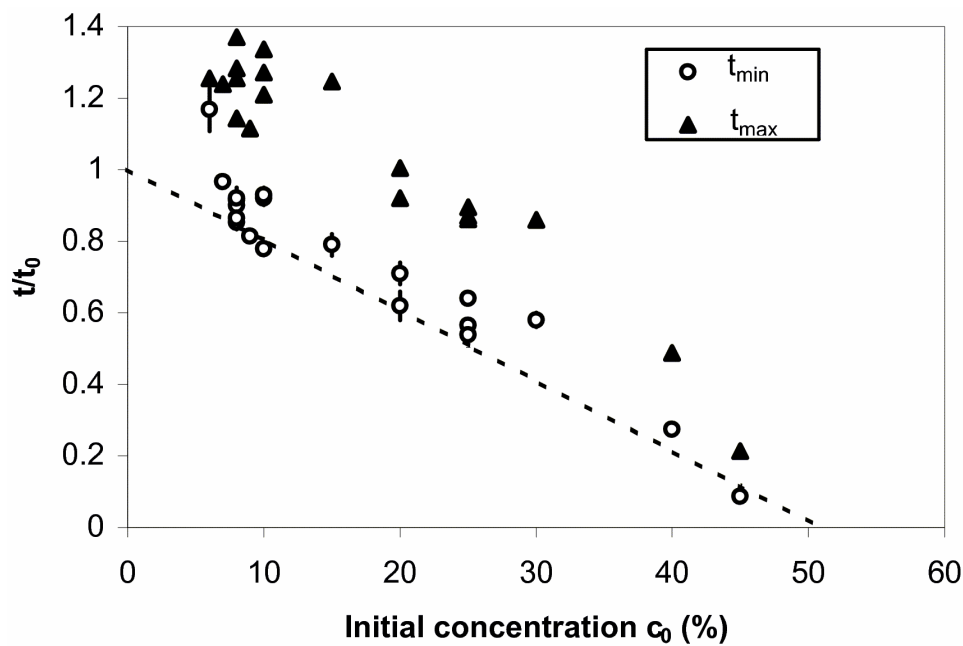

**Fig.4** Normalised values of $t_{min}$ and $t_{max}$, the time when the droplet height reaches a minimum and maximum respectively. Error bars are due to uncertainties in the exact time of the extrema and in the extrapolation to determine the normalising value $t_0$. The straight line fit through the $t_{min}$ data has y-intercept fixed at 1 and gives an x-intercept of 50% (Eq.3).

In Fig. 4 we plot the variation of normalised $t_{min}$ and $t_{max}$ with initial concentration $c_0$. The vertical error bars combine uncertainties in the exact time at which the extrema occur and in the extrapolated value of $t_0$. Both the $t_{min}$ and $t_{max}$ data show a steady decrease as $c_0$ increases.

Fig.5 shows the $c_0$ dependency of normalised $h_{min}$ and $h_{max}$ values. Both values increase with concentration but in different manners: $h_{min}$ shows a smooth increase above 3%; $h_{max}$ has a steep initial increase, rising from 0 at $c_0 = 2\%$, to over 1 at $c_0 = 8\%$, and then remaining roughly constant at 1.35.

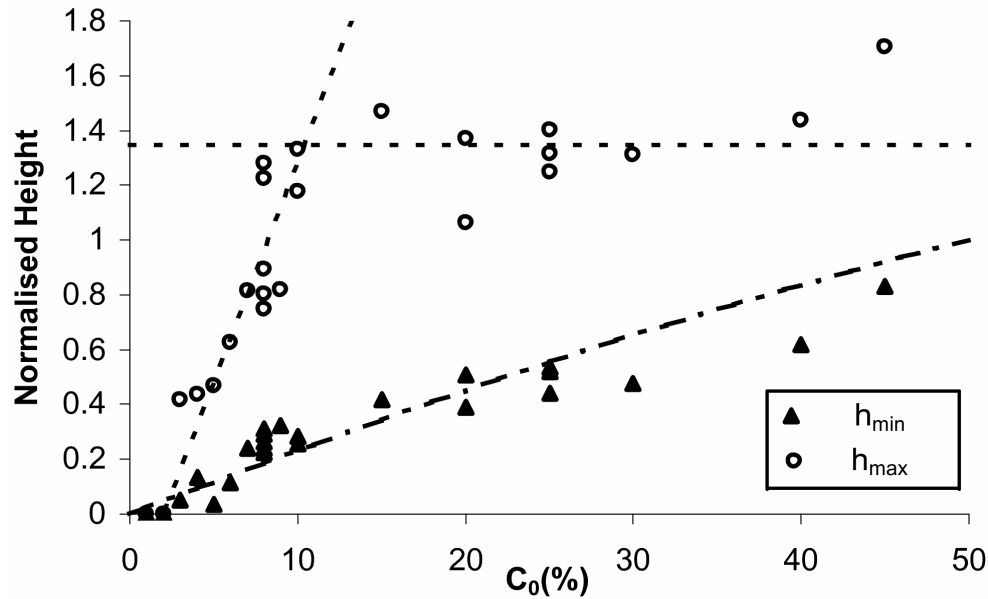

**Fig.5** Normalised minimum and maximum heights for a range of initial concentration values $c_0$. The dash-dot line is a prediction for $h_{min}$ using previously determined value of $c_{min}=50\%$ (Eq.7). The straight dashed lines are guides to the eye for $h_{max}$ values.

Finally, we use the volume data at $t_{min}$ and $t_{max}$ to calculate the overall droplet concentrations at these times using

$$c_{min/max} = \frac{c_0 V_0}{V_{min/max}} \, . \tag{2}$$

In Fig.6 we plot $c_{min}$ and $c_{max}$ values which show no clear dependency on $c_0$. Values of $c_{min}$ have an average of $\overline{c_{min}} = 49 \pm 8\%$ and values of $c_{max}$, despite greater uncertainties due to difficulties in determining volume at later times, are less scattered and have an average of $\overline{c_{max}} = 73 \pm 6\%$.

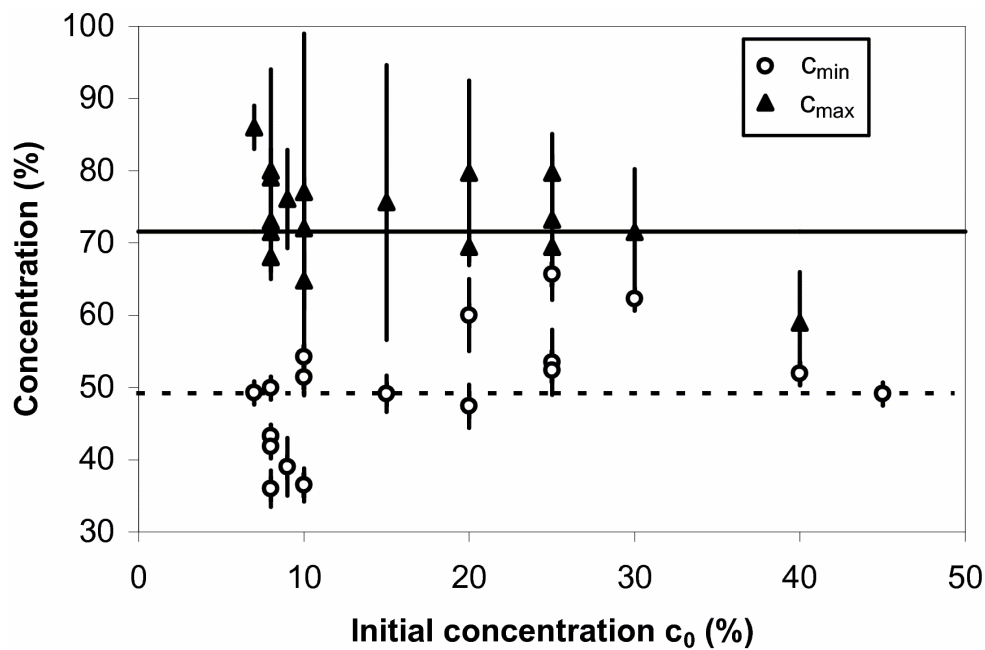

**Fig.6** Droplet concentration when the droplet height is a minimum ($c_{min}$) and a maxima ($c_{max}$). $c_{min}$ values are scattered around an average of 49±8% which is in agreement with the reported saturation concentration of PEO solutions $c_{sat} \approx 50\%$ [19]. $c_{max}$ values, despite greater uncertainties due to difficulties in determining volume precisely, are scattered around an average of 73±6%.

## Analysis and Discussion

### Buckling Skin Model

The time course of the droplet height plotted in Fig.2 and for all other samples (a slow initial decrease followed by a rapid increase) is similar to published measurements on dextran[10], which are well explained by the model of a buckling skin with constant surface area. However, this mechanism does not agree with our observations. Firstly, our data for the temporal evolution of $A/A_0$ presented in Fig.3 show, for all concentrations, a noticeable increase in surface area during the growth phase, even within our significant uncertainties. Secondly, PEO is known to crystallise into spherulites at high concentrations[17], rather than form an amorphous glass. Thirdly, a glassy skin covering the droplet would prevent the growing deposits from falling over during drying, in contradiction with what is seen in our "failed" experiments.

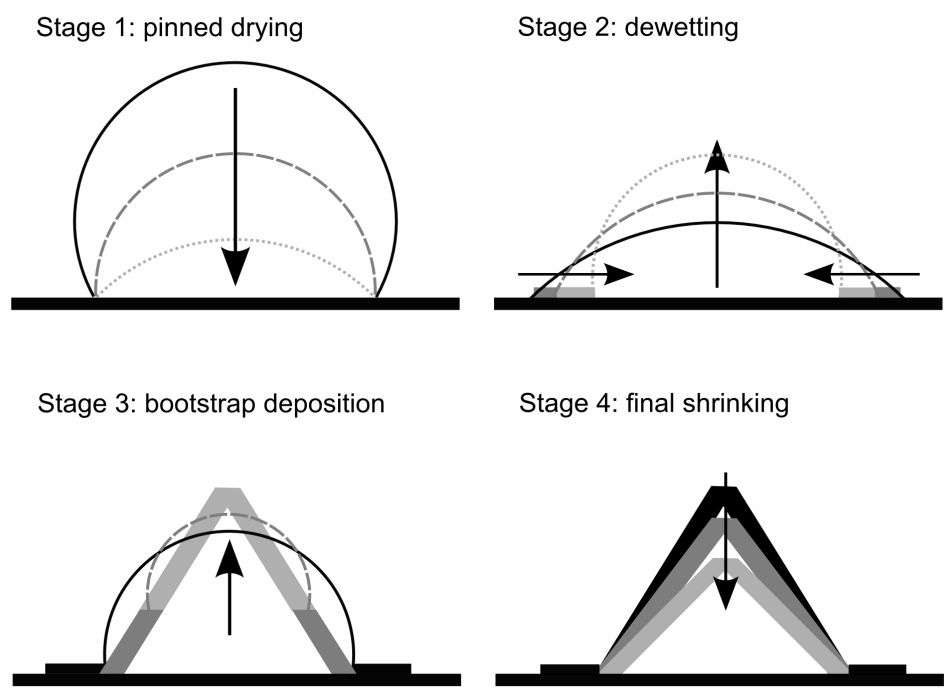

**Fig.7** Schematic drawing of the proposed four drying stages. Thin lines indicate liquid surfaces, thick regions represent solid deposits. Progress within each stage is from solid black to dashed dark grey to dotted light grey.

**Four-stage Deposition Model**

To understand the observed drying process, we develop an alternative model, in which we identify four distinct drying stages, including a novel "bootstrap" stage. The model is described below and graphically in Fig.7, followed by discussion of specific predictions.

**Stage 1.** During Stage 1 the droplet shows typical pinned contact line drying behaviour with a constant droplet radius r. To accommodate the reducing droplet volume, h and $\theta$ both decrease but typically $\theta$ remains above the receding contact angle (measured in separate experiments to be around 5° for $c_0 = 15\%$). The evaporation rate is greatest at the contact line (provided $\theta < 90°$), and is sustained by solvent within the droplet flowing radially outwards[7] (evidence of which is provided in Video 5 in the ESI[†]). When the droplet concentration reaches saturation $c_{sat}$, semi-crystalline spherulites are precipitated, in which the water molecules are tightly bound through hydrogen-bonding to the polymer[17] so are not able to participate in outward flow. Consequently, the contact line can not remain pinned and must retract. At this time the droplet height reaches its minimum value $h_{min}$ with concentration $c_{min}$.

**Stage 2.** During Stage 2 the contact line of the remaining liquid, initially a flatish puddle, retracts, driven by the large difference between the actual contact angle and the equilibrium contact angle: the droplet in Fig.1 with $c_0 = 40\%$, approaching $c_{sat}$ has $\theta \sim 90°$, but at the start of Stage 2 when the concentration for all droplets is also $c_{sat}$, $\theta$ is much lower due to pinning in Stage 1. This is a dewetting transition and as shown in Fig.2, there is an observed decrease in r and increase in surface area. Although an increase in A will result in a corresponding increase in surface energy, measurements to quantify the surface tension lowering properties of PEO[18] show a maximal reduction for polymers with molecular weight of 80kDa, close to those used here. Provided evaporation is slow compared to the speed of the retracting contact line, h will increase rapidly, reminiscent of behaviour seen with liquid mixtures[12]. The receding contact line leaves behind a thin layer of dry polymer, similar to the gelled foot reported in previous studies of dense particle suspensions[9]. Stage 2 finishes when $\theta$ reaches a value around $\theta \approx 80°$ and the contact line stops retreating.

**Stage 3.** As shown in Fig.2 and observed for other samples, h continues to increase even though the radius of the deposit r remains constant. In fact, the observed kink in the h data is another indicator for the transition between Stage 2 and 3. During Stage 3 the liquid droplet, at concentration $c_{sat}$, coexists with solid spherulites at $c_{spher}$. Continuing evaporation, via constant contact angle mode[19], leads to a diminishing liquid phase in place of further spherulites, which are deposited in a ring at the contact line. The remaining liquid droplet is fenced in and squeezed upwards by the growing deposit, as illustrated in Fig.8. We call this process "bootstrap building" to encapsulate how the droplet seems to push itself upwards. Fig.8 shows snapshots of this process for a droplet with $c_0=10\%$ and ESI† Video 2 for $c_0=8\%$. Eventually, the liquid droplet is entirely supported by the deposit and loses contact with the substrate leaving behind a solid structure, which when viewed from underneath (ESI Video 5) or carefully cut open, is seen to be partially hollow. Stage 3 ends when all liquid phase has precipitated as spherulites and the overall droplet concentration is $c_{spher}$. At this time $t_{max}$ the overall structure reaches its maximum height $h_{max}$. For the sake of clarifying the distinctions between the stages, we ignore the effects of evaporation during Stage 2; in practice there is nearly always overlap of Stages 2 and 3.

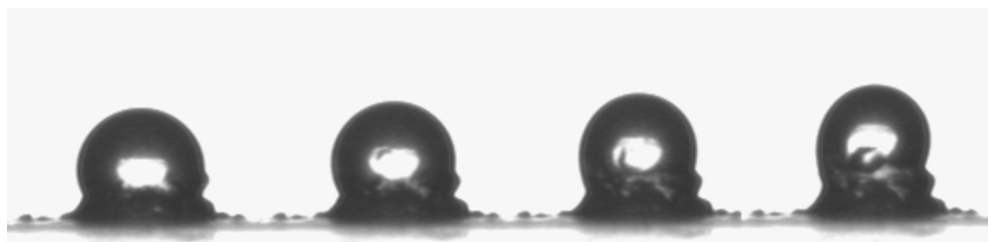

**Fig.8** Time-sequence taken during stage 3 ($c_0$=10%) showing a liquid droplet being raised by the solid deposit. Time between images is 20s.

**Stage 4.** During Stage 4, the solid structure formed at the end of Stage 3 shrinks slowly by up to 10% in height as the remaining water within the spherulites evaporates. For larger initial concentrations ($c_0 > 30\%$) a small amount of liquid can be trapped inside the solid cone which is then forced through the top by the shrinking structure, resulting in the eruption seen at time t=12000s in Fig.1 for $c_0 = 30\%$ and 40% and in ESI[†] Video 4. Stage 4 ends when the droplet is completely dry. During this stage, the forces generated by the shrinking structure stuck to the coverslip can be sufficiently strong to cause the glass coverslip to bend upwards[20].

**Predictions of Four-stage model**

The model presented above lends itself to various experimental verifications. Details of such tests are discussed.

**Values at minimum and maximum height.** The model allows prediction of the value of several parameters (including time, concentration and height) when the droplet reaches minimum and maximum height. Firstly, it predicts that the minimum height should occur at the same concentration for all $c_0$, at $c_{sat}$. Fig.6 shows the measured average value to be $\overline{c_{min}} = 49 \pm 8\%$ , in agreement with the literature value of $c_{sat} \approx 50\%$[16]. The model also predicts that the concentration at the maximum height should be independent of $c_0$ and occur at $c_{spher}$, which is also confirmed in Fig.6 where the value is calculated as $\overline{c_{max}} = 73 \pm 6\%$ .

Secondly, the normalised values of $t_{min}$ can be calculated by first integrating Eq.1 to find V(t) (assuming volume loss continues at its initial rate which from Fig.2 seems reasonable), and then by combining with Eq.2 to give

$$\frac{t_{min}}{t_0} = 1 - \frac{c_0}{c_{min}}.$$

(3)

Fig.4 shows that this equation gives a good fit to the $t_{min}$ data and provides a consistent estimate of $c_{min} = c_{sat}$ as the intercept on the x-axis at 50%. Applying a similar analysis to the $t_{max}$ data is more complicated, as the assumption regarding rate of evaporation is no longer valid.

Finally we can find $h_{min}(c_0)$, provided we make the additional assumption that the droplet has the shape of a spherical cap, so its volume can be written as

$$V = \frac{1}{6}\pi r^3 \left( X^3 + 3X \right) \tag{4}$$

in terms of base radius r and the ratio X

$$X = \frac{h}{r} = \tan\frac{\theta}{2} . \tag{5}$$

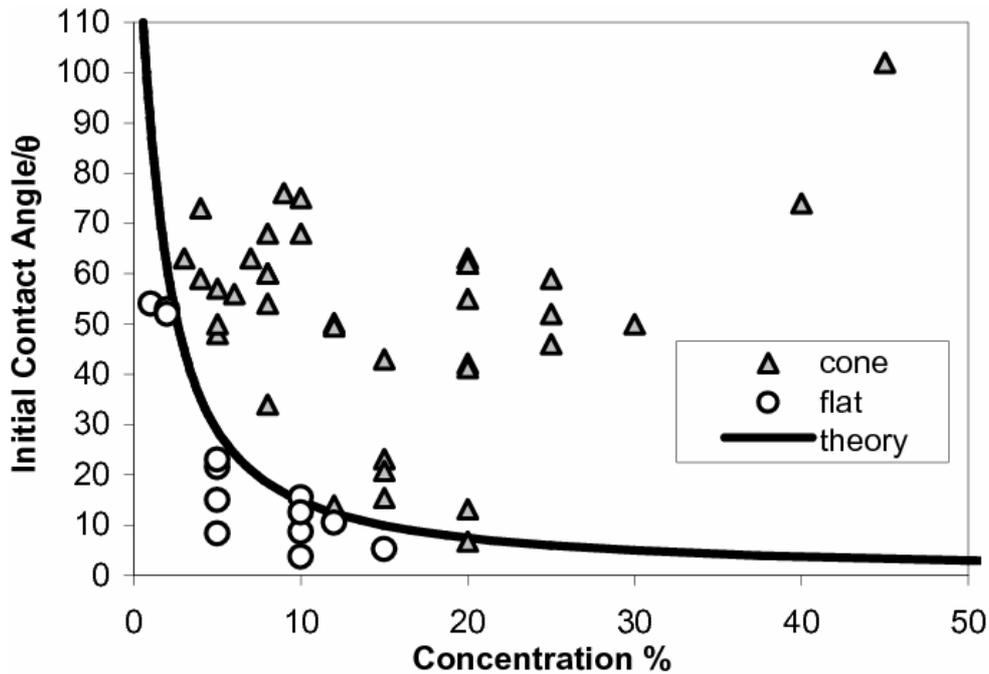

**Fig.9** "Phase diagram" depicting whether or not a given sample, characterised by its initial contact angle and initial concentration, is observed to form a central solid deposit. The theoretical curve (Eq.8) separates two behaviours: above and to the right are samples in which the concentration reaches $c_{sat} \approx 50\%$ first; for samples below and to the left, the contact angle reaches the receding value, $\theta_r \approx 3°$ first.

We then combine Eq.2 and 5 to write the unknown $X_{min}$ in terms of known parameters

$$X_{min}^3 + 3X_{min} = \frac{c_0}{c_{sat}} \left( X_0^3 + 3X_0 \right) = 2D , \tag{6}$$

in which the r terms cancel as the droplet is pinned during Stage 1 and D is a constant. This depressed cubic can be solved for $X_{min}$, and then normalised by $X_0$ to give the analytical expression for $h_{min}(c_0)$:

$$\frac{h_{\min}}{h_0} = \frac{X_{\min}}{X_0} = \frac{1}{X_0}\left(\frac{1}{\sqrt[3]{-D+\sqrt{D^2+1}}} - \sqrt[3]{-D+\sqrt{D^2+1}}\right) \qquad (7)$$

Taking an initial contact angle of 70° and $c_{sat} = 50\%$, we obtain the curve in Fig.5 which agrees well with the experimental data. We are currently working on the theory to predict the behaviour of $h_{max}(c_0)$.

**Nature of the deposit.** We propose a simple argument to predict whether a given droplet will form a tall solid deposit or a flat disk. For most droplets studied, depinning at the end of Stage 1 occurs due to the concentration of the droplet reaching $c_{sat}$ from which point the droplet proceeds to Stage 2 and forms a tall central deposit. However, there is an alternative scenario: a droplet will also depin when the contact angle falls below the receding contact angle $\theta_r$,[8] measured to be around 5° for a droplet with concentration of 15%. In this case, the droplet concentration will be less than $c_{sat}$ and Stage 2 does not take place. The contact line retracts as the concentration increases, but bootstrap building does not occur. Using Eq.5 to write X in terms of $\theta$ we define a critical concentration, dependent on $\theta_0$ and $\theta_r$ only:

$$c_{crit} = c_{sat}\left(\frac{X_r^3 + 3X_r}{X_0^3 + 3X_0}\right). \qquad (8)$$

For $c_0 > c_{crit}$ the droplet forms a tall deposit; for $c_0 < c_{crit}$ the droplet forms a thin deposit. To test this prediction, we deliberately prepared droplets with low $\theta_0$ by pipetting a large droplet and then removing much of the liquid. These samples were analysed before and after drying to measure $\theta_0$ and to check whether they formed a conical central deposit or not. The results from all previous experiments and these additional low $\theta_0$ samples are plotted on Fig.9. The curve is a plot of Eq.8 using $c_{sat}=50\%$ and $\theta_r=3°$ and shows good agreement with the experimental observations.

## Conclusions

From our experimental study of drying droplets of aqueous poly(ethylene oxide) solutions we conclude that the shape of the final solid deposit (either tall, conical and often hollow, or flat and circular) depends sensitively on both the initial droplet concentration and the initial contact angle. Despite superficial similarities with previous studies using dextran solutions[10], in which the deposit shapes were attributed to buckling of an incompressible glassy polymer skin, we demonstrate that a different mechanism must be at work here as the surface area consistently increases during the growth phase.

To rationalise our observations, we propose a four stage drying and deposition process, including a novel bootstrap stage during which the liquid droplet is lifted up on freshly precipitated solid. We argue that droplets reach a minimum height when they first begin to precipitate solid spherulites, when their concentration reaches a saturation value, determined here to be in good agreement with the literature value of 50%. We propose that whether a given droplet forms conical central deposits or not is controlled by which occurs first: the concentration reaching saturation or the contact angle dropping below the receding contact angle. This criterion agrees well with our observations. As PEO is a common laboratory polymer with varied industrial applications (e.g. as a food additive[21], in the preservation of wooden artefacts[22] and in protein crystallisation[23]), understanding its drying behaviour could have practical or technological implications.

## Acknowledgements

DW is funded by EPSRC/Unilever (EP/C547462/1), KB by the Nottingham Trent University Vice Chancellor's Bursary Scheme and DF partially by the Royal Society and the Nuffield Foundation. We thank Prof. Glen McHale, Dr. Mike Newton and Dr. Neil Shirtcliffe for discussions and use of their DSA hardware and software.

## Notes and references

[‡]This article is part of a virtual themed issue containing papers from the 14th UK Polymer Colloids Forum Annual Meeting, held in Hull, UK, in September 2009.

[†]**Electronic Supplementary Information** (ESI) available on www.scivee.tv

Videos 1 to 4 show typical drying behaviour of different concentration droplets. The duration of each experiment was approximately 2 hours, and the size of each droplet initially around 75µl and the distance across the image approximately 10mm. Timings given in the captions below are given relative to the video files, to aid identification of the various stages and processes. It is also useful, if viewing the files in Quicktime (ver 7.5) to use the Jog/Shuttle control in A/V controls.

**Video 1** (http://www.scivee.tv/node/16842) shows drying of a sample with $c_0 = 5\%$ and exhibits pinned drying (stage 1) for around the first 5 seconds, before depinning when the contact angle becomes lower than the measured receding contact angle, $\theta_r$. Stage 2 and bootstrap building do not take place.

**Video 2** (http://www.scivee.tv/node/16843) shows drying of a sample with $c_0 = 8\%$. This exhibits pinned drying (stage 1) for around the first 5 seconds, at which point the contact line depins and the droplet undergoes a dewetting transition (stage 2) until 9 seconds. At this time the contact angle remains constant at just over 90 degrees, and the droplet climbs on top of the solid deposit, bootstrap building (stage 3). At 11 seconds the outer surface appears completely solid and the maximum height is reached. Stage 4 drying follows as the solid shape slowly shrinks. The base radius of the final cone is less than a quarter of that of the initial droplet.

**Video 3** (http://www.scivee.tv/node/16846) shows drying of a droplet with $c_0$ = 25% and again exhibits stage 1 pinned drying for the first 5 seconds. However, the solid deposit is already much thicker for this droplet. As the contact angle is increasing (stage 2) deposition occurs simultaneously (stage 3). The maximum height is reached at 10 seconds from which point, late stage drying (stage 4) accounts for the slow decrease in height. The final cone has half the base radius of the initial droplet.

**Video 4** (http://www.scivee.tv/node/16847) shows drying of a droplet with $c_0$ = 40% and has a very short stage 1, around 1 second. The contact angle increases until 3 seconds and bootstrap deposition (stage 3) continues until 4 seconds. However, there is still liquid inside the structure at this point, which is ejected as stage 4 begins. The base radius of the final cone is around three quarters that of the initial droplet.

**Video 5** (http://www.scivee.tv/node/16848) was recorded using an inverted microscope (Nikon Eclipse TE2000-S) with 2× objective lens. The images measure 5mm across and the frame rate is increased by a factor of 300. The bright specks in the droplet are small clusters of polymer that would not dissolve, and help to visualise the flow within the droplet. Pinned drying (stage 1) occurs for the first 5 seconds, during which there is clear evidence for recirculation flow at the contact line, with the liquid near the base flowing radially outwards and moving inwards above. As the flow at the edge ceases, a bright region of solid deposit appears behind the retreating liquid droplet (stage 2). There is no longer evidence of recirculation flow within the droplet. At 14 seconds, the deposit has reached its maximum height at the end of stage 3 and final drying begins. A wide solid ring has been deposited with a liquid region in the centre. From around 18 seconds, darker lines appear in the bright deposit, indicating completely dry areas and at 24 seconds, the central liquid region begins to dry (darken) and by 40 seconds, the structure is hollow.